\documentclass[doublecol,figures,final]{epl2}

\usepackage{exscale}

\usepackage[intlimits]{amsmath}

\usepackage{textcomp}

\usepackage{xspace}

\usepackage{units}

\usepackage{stackrel}

\usepackage{paralist}

\newcommand \expb[1] {\ensuremath{\exp{\left(#1\right)}}}                   
\newcommand \avg [1] {\ensuremath{\left<{#1}\right>}}                       

\newcommand {\kBT} {\ensuremath{k_{\mathrm{B}}T}\xspace}              

\newcommand {\ie} {\textit{i.e.}\xspace}			
\newcommand {\eg} {\textit{e.g.}\xspace}			
\newcommand {\ea} {\textit{et\,al.}\xspace}		

\newcommand \eqnref [1] {\mbox{eq.~(\ref{#1})}}		
\newcommand \figref [1] {\mbox{fig.~\ref{#1}}}		

\usepackage{amsmath}
\title{Single polymer adsorption in shear: flattening versus hydrodynamic lift and
surface potential corrugation effects}

\shorttitle{Single polymer adsorption in shear} 

\author{A.~Serr, C. Sendner, F. M\"uller, T.~R.~Einert  \and R.~R.~Netz}
\institute{Physics Department, TU Munich - 85748 Garching, Germany}

\date{\today}

\pacs{82.35.Gh}{Polymers on surfaces; adhesion}
\pacs{83.50.-v}{Deformation and flow}
\pacs{82.37.Gk}{STM and AFM manipulations of a single molecule}

\abstract{The adsorption of a single polymer to a flat surface in shear is investigated
  using Brownian hydrodynamics simulations and scaling arguments.  Competing effects are
  disentangled: in the absence of hydrodynamic interactions, shear drag flattens the chain
  and thus enhances adsorption.  Hydrodynamic lift  on the other hand gives rise to
  long-ranged repulsion from the surface  which preempts the surface-adsorbed state via a discontinuous
  desorption transition, in agreement with theoretical arguments.  
Chain flattening is dominated by hydrodynamic lift, so overall,
shear flow weakens the adsorption of flexible polymers. 
Surface friction due to  small-wavelength  surface potential corrugations
is argued to weaken the surface attraction as well.
}

\begin{document}
\renewcommand{\revision}{}
\maketitle

The adsorption of polymers on surfaces is at the base of  many applications such as 
surface modification, colloidal stabilization and flocculation. 
By tailoring adsorption properties, polymer additives can be used as lubricants, 
adhesives, strength enhancing agents or coagulants. 
Many experimental and theoretical investigations on the basic research level exist, see  \cite{Netz03} for references. 
The theoretical foundation for polymer adsorption has been laid down by de Gennes \cite{deGennes81}. 
Recent works fine-tuned the microscopic picture and addressed the adsorption of charged or stiff 
polymers\cite{Fleer,Yamakov99}.

Although it was realized early that non-equilibrium effects are very relevant for polymer
adsorption because of the long relaxation times\cite{deGennes81}, most theoretical and experimental
work concentrated on
equilibrium aspects\cite{Netz03,Eisenriegler82}. 
However,
technological applications involving polymer adsorption are typically far from  equilibrium. 
Along the same lines, many biological mechanisms involve adsorption of macromolecules
in shear flow, e.g. the initiation of the coagulation cascade involving the van-Willebrand-factor\cite{Schneider}
or the adsorption of E. coli  on surfaces\cite{Vogel}.
\revision{Non-equilibrium aspects of polymer adsorption  receive growing attention both from the 
experimental \cite{Lee85,McGlinn88,Besio,Chin91,Chang,Nguyen,Kim07,He1} 
and theoretical point of view \cite{Milchev96,Manias95,Chopra,Panwar05,Kumar2,Kumar3,He2} but 
are still less well understood than the equilibrium case. 
In the context of the current theoretical investigation, two opposing effects are relevant:
A  polymer that is either pulled laterally by a terminally exerted 
force\cite{serr} or subject to shear\cite{He1,He2} at an adsorbing surface 
is flattened. In the absence
of hydrodynamic interactions, this has been found to {\em enhance} adsorption in simulations\cite{Manias95,Panwar05}.
On the other hand, hydrodynamic effects generate a shear-dependent 
 long-ranged lift force that 
decays quadratically with distance from the surface for
dumbbell models\cite{Graham} as well as 
 for stiff and flexible  polymers\cite{Ladd,sendner}
 which in simulations has been found to {\em weaken} 
 polymer adsorption\cite{Kumar2}.
In the present work we disentangle those opposing aspects and 
consider flexible polymer at adsorbing surfaces in shear flow using simulations and theory:
In the absence of hydrodynamics,  we rationalize the shear-induced flattening of the chain and 
the adsorption enhancement observed in simulations by linear-response theory.
As we turn on hydrodynamic interactions in simulations, long-ranged hydrodynamic lift effects
dominate over chain flattening effects and lead to  chain desorption via a discontinuous 
transitions, as predicted by scaling arguments.
An analysis of a single-particle model   
based on the Fokker-Planck equation suggests that lateral motion over a surface
with an inhomogeneous or corrugated potential
 weakens the effective attraction, in line with previous simulations\cite{Kumar3}
Our overall conclusion is that shear flow weakens the  adsorption of flexible
polymer chains on planar surfaces.}


  \begin{figure*}
    \begin{minipage}[t]{.32\textwidth}
      \includegraphics[width=\textwidth]{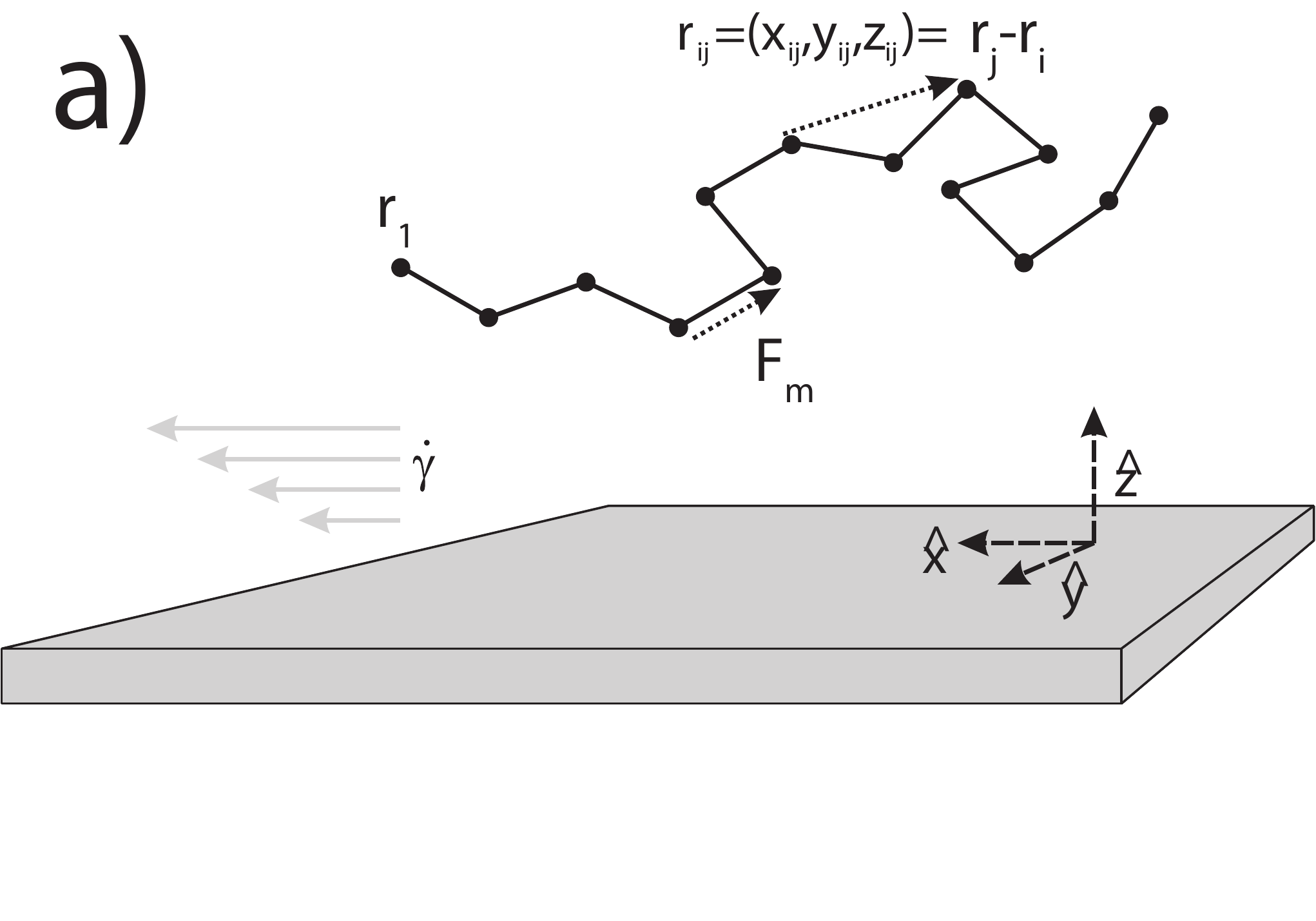}
    \end{minipage}%
    \hfill%
    \begin{minipage}[t]{.32\textwidth}
      \includegraphics[width=\textwidth]{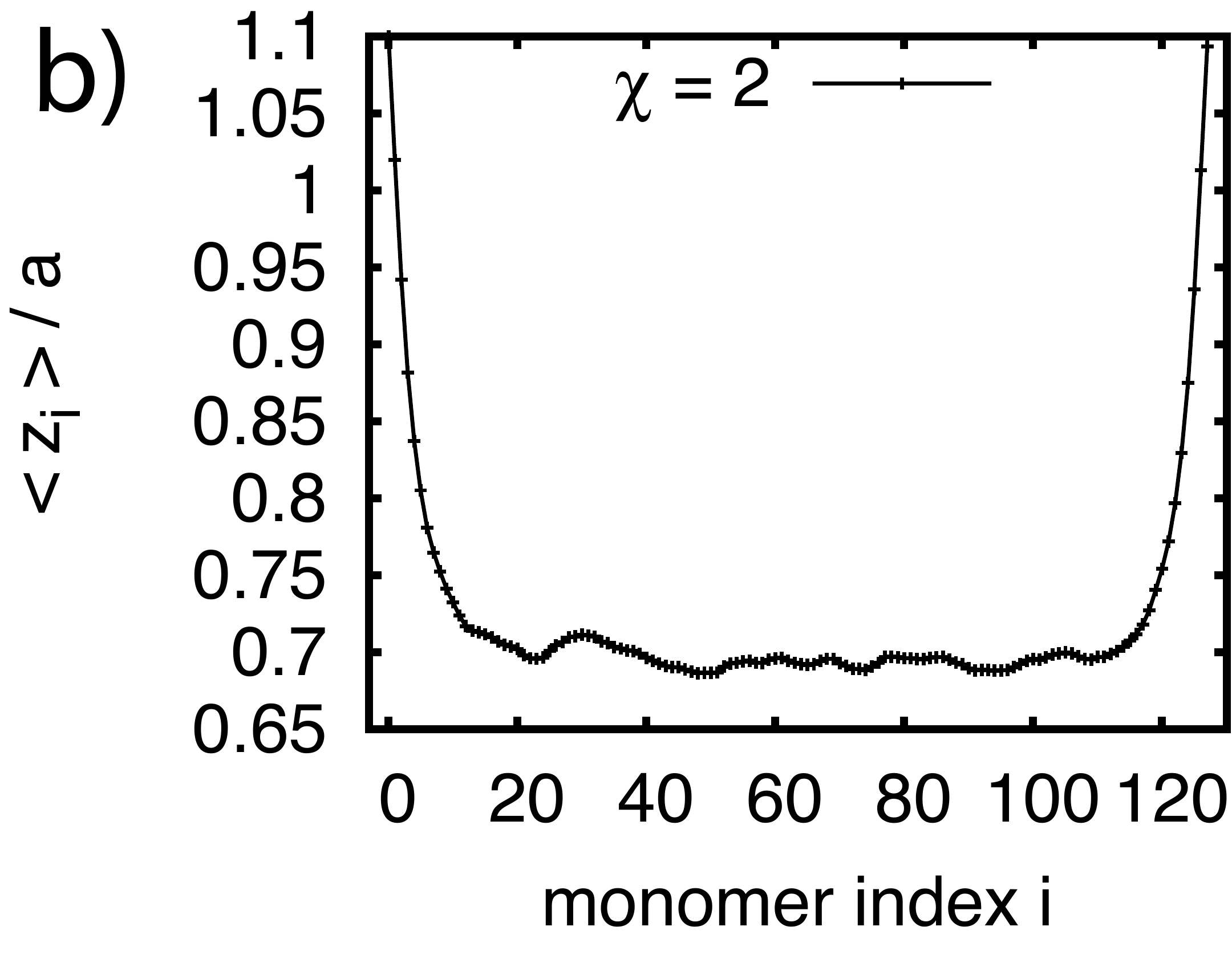}
    \end{minipage}%
    \hfill%
    \begin{minipage}[t]{.32\textwidth}
      \includegraphics[width=\textwidth]{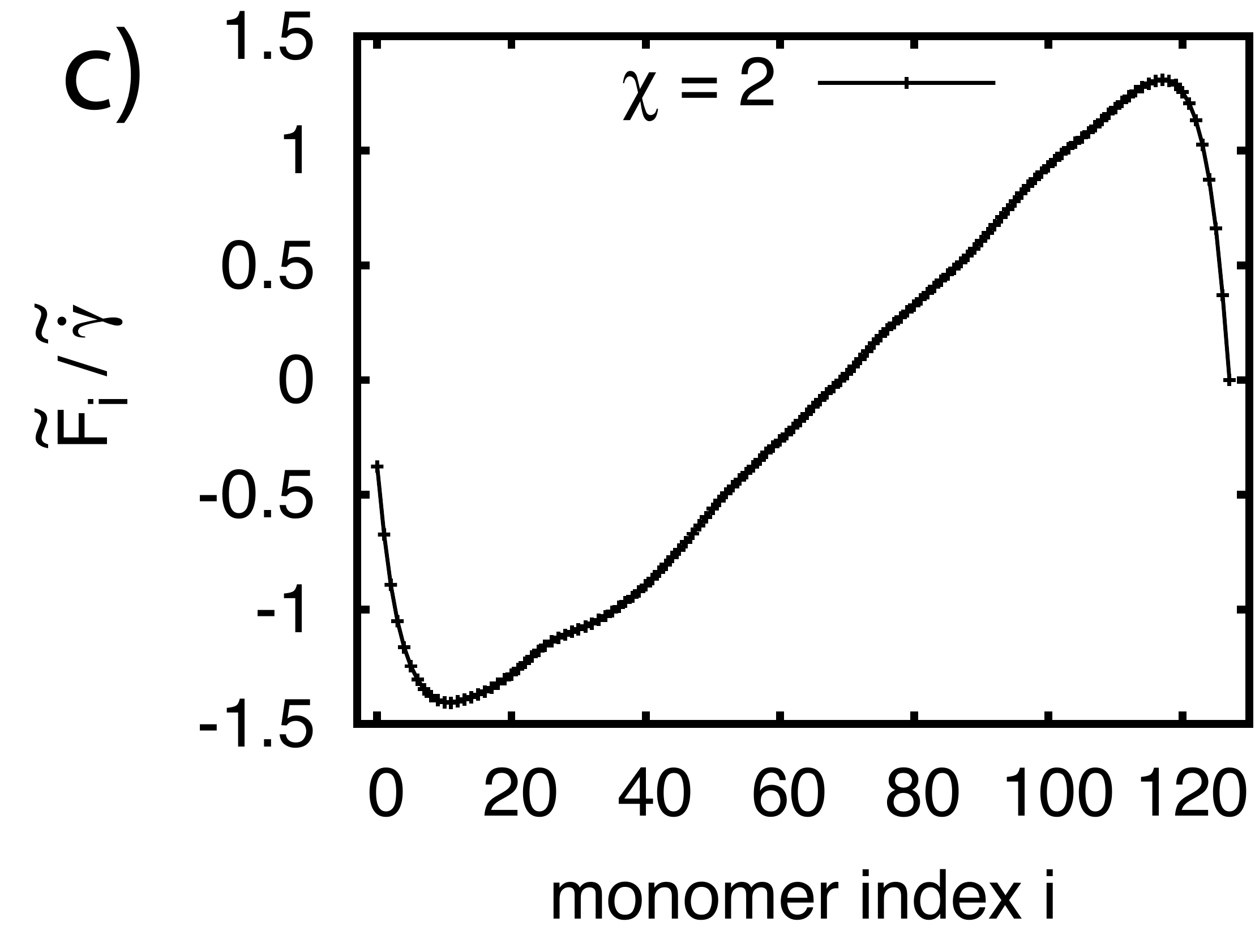}
    \end{minipage}%
    \caption{a) Schematic picture of the polymer model for non-equilibrium adsorption in
      shear.  The $\hat{z}$ axis coincides with the surface normal. The position of
      monomer $i$ is denoted as $\bm{r}_i=(x_i,y_i,z_i)$, the distance between two
      monomers $i$ and $j$ as $\bm{r}_{ij}=(x_{ij},y_{ij},z_{ij})= \bm{r}_j -
      \bm{r}_i$. The force $\bm{F}_m $ is the force exerted by monomer $m+1$ on monomer
      $m$.  A simple shear flow in $\hat{x}$ direction acts with shear rate
      $\dot{\gamma}$.  b) The mean height profile $\{z_i /a \}$, $i = 1\ldots 128 $ as
      a function of the monomer index $i$ for a 128mer obtained from equilibrium BD
      simulations at the adsorption strength $\chi=2$ and decay length $\tilde{\kappa}^{-1} =
      1/2$.  c) The mean rescaled and averaged lateral force
      $\tilde{F}_i/\dot{\tilde{\gamma} }= \langle F_i^{(x)} \rangle \mu_0 / \dot{\gamma}
      a$ as a function of the bond index $i$ for a 128mer.  The profiles are obtained from
      the height distributions shown in b) via eq.~(\ref{force}). }
    \label{fig1}
  \end{figure*}

In the simulations, a polymer is composed of $N$ spherical beads with diameter $a$. 
Neglecting particle inertia, the time evolution 
of the position of monomer $i$ obeys the Langevin equation~\cite{doi}
\begin{multline}
  \dot{{\bf r}}_i(t) = - \sum_{j=1}^N \bm{\mu}_{ij} \bm{\nabla}_{r_j} U({{\bf       \{r}_N\}}) +   \\
k_B T  \sum_j \bm{\nabla}_{r_j} \bm{\mu}_{ij}  + \dot{\gamma} z_i   \bm{\hat{e}}_x + \bm{\xi}_i(t),
  \label{langevin}
\end{multline}
a typical setup of the simulation is shown in fig. \ref{fig1}(a).
Hydrodynamic effects are incorporated via the mobility matrix $\bm{\mu}_{ij}$ which is obtained from the 
Green's function $  {\bf G}^W({\bf r}_i,{\bf r}_j)$
of the Stokes equation that satisfies the no-slip condition on a planar boundary~\cite{blake,sendner}.
Performing a multipole expansion to second order in terms of the bead diameter $a$,
\revision{which is accurate for bead-bead and bead-surface separations larger than $ \approx 2a$, } we write~\cite{Kim}
\begin{equation}
  \boldsymbol{\mu}_{i j}({\bf r}_i,{\bf r}_j) =
  \left[ 1 + \frac{a^2}{ 24} \boldsymbol{\nabla}^2_{{\bf r}_i} \right] 
  \left[ 1 + \frac{a^2}{ 24} \boldsymbol{\nabla}^2_{{\bf r}_j} \right] {\bf G}^W({\bf     r}_i, {\bf r}_j) 
  \label{multipole}
\end{equation}
for $i \not= j$.
The self mobility tensor $\boldsymbol{\mu}_{ii}$ is obtained in the limit ${\bf r}_i \to {\bf r}_j$ and 
regularized far away from the surface and set equal to the bulk sphere mobility $\mu_0 = 1/(3 \pi \eta a)$~\cite{Kim}.  
\revision{ This hydrodynamic treatment is valid in the incompressible low-Reynolds-number limit and assumes 
instantaneous propagation of hydrodynamic effects, which is accurate for most experimentally relevant 
situations. Effects due to the discreteness of water molecules are in fact negligible for all but
sub-nanometer length scales\cite{sendner2}.}
In the free-draining simulations, the mobility tensor is set to be diagonal and constant in space,
\begin{equation} \label{FD}
  \boldsymbol{\mu}_{i j}({\bf r}_i,{\bf r}_j) = \delta_{ij} \mu_0  \boldsymbol{1}
\end{equation}
where $ \boldsymbol{1}$ is the diagonal unitary tensor.
The externally imposed flow is a linear shear with shear rate $\dot{\gamma}$.
For the Brownian dynamics simulations, we use the rescaled, discrete version   \cite{ERMAK} 
of eq.~(\ref{langevin}),
  \begin{multline}
    \label{eq:5}
    {\bf \tilde{ r}}_i(t+\Delta t) - {\bf \tilde{ r}}_i( t) = -     \sum_{j=1}^N \bm{\tilde{\mu}}_{ij} \bm{\nabla}_{\tilde{r}_j} u({\{ {\bf         \tilde{{r}}}_N\}}) \\ + \sum_j \bm{\nabla}_{\tilde{r}_j}     \tilde{\bm{\mu}}_{ij} + \tilde{\mu}_0 \dot{\tilde{\gamma}} \tilde{z}_i     \bm{\hat{e}}_x + \bm{\tilde{\xi}}_i(t).
  \end{multline}
\begin{figure*}
   \begin{minipage}[t]{.32\textwidth}
      \includegraphics[width=\textwidth]{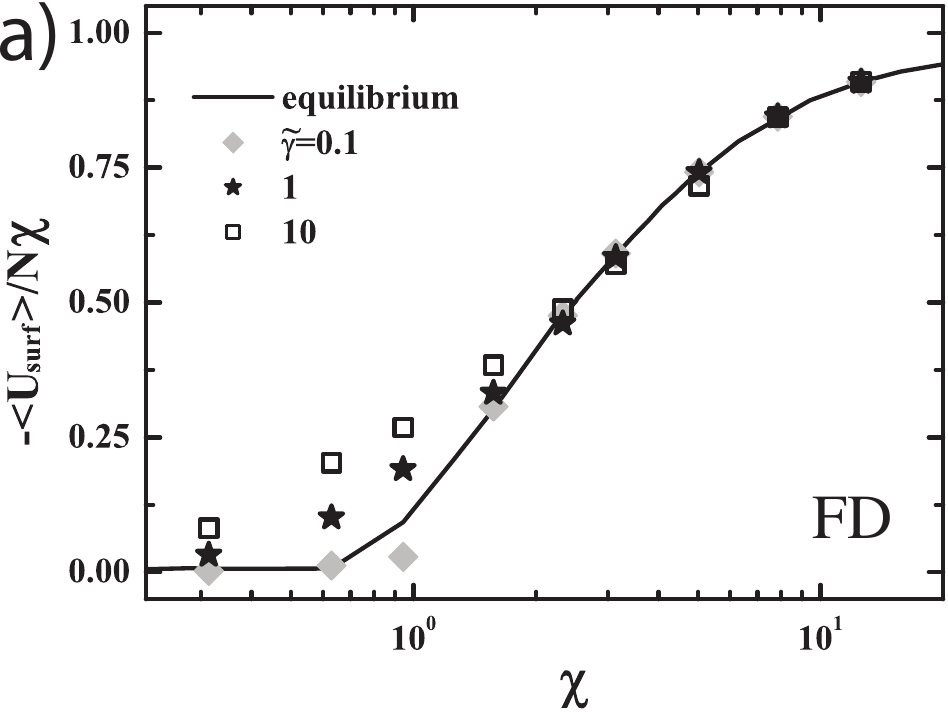}
    \end{minipage}
    \hfill
    \begin{minipage}[t]{.32\textwidth}
      \includegraphics[width=\textwidth]{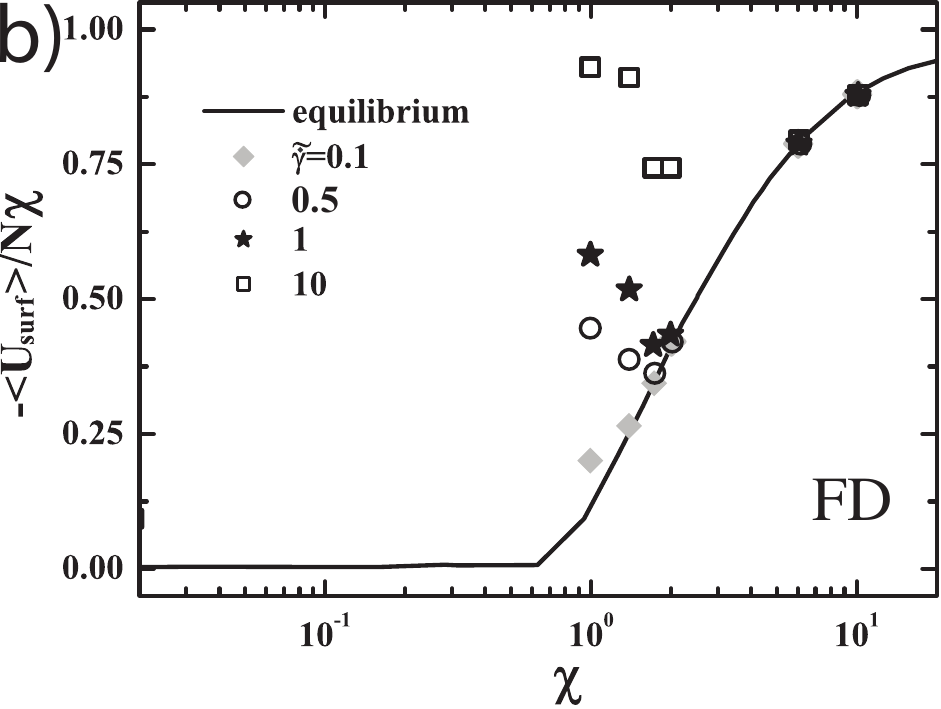}
    \end{minipage}
   \hfill
    \begin{minipage}[t]{.32\textwidth}
      \includegraphics[width=\textwidth]{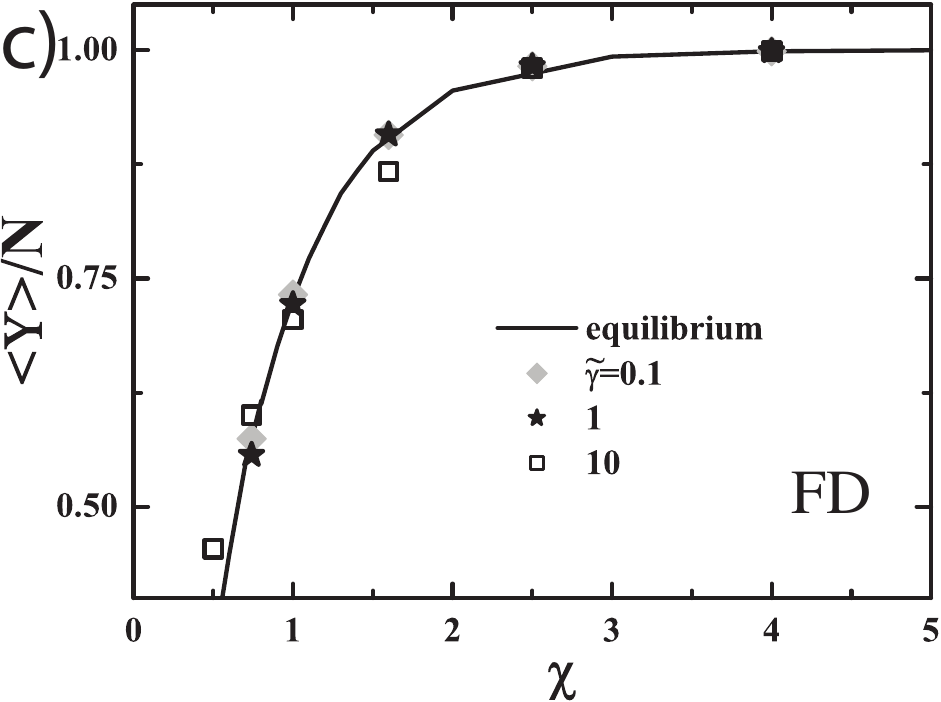}
    \end{minipage}
    \caption{ Free draining (FD)  results.
    a) Simulation results for the mean normalized adsorption potential $-\langle u_\mathrm{surf} \rangle /N\chi $ 
 for a 128mer as a function of the surface interaction parameter $\chi$ at various shear rates  
 $\dot{\tilde{\gamma} }$.
 b) Results from the linear-response theory in which the lateral stretching force distribution 
 due to shear is calculated from the equilibrium monomer height distribution via eq.~(\ref{force}) and
 used to derive the effective adsorption strength via eq.~(\ref{chieff}), 
 see text for details. The linear response theory  captures the shear-induced adsorption enhancement  only qualitatively.
 c) Fraction of adsorbed monomers $\langle \Upsilon \rangle /N$ from simulations.
 A monomer $i$ is defined adsorbed if $\tilde{z}_i < \tilde{\kappa}^{-1}$.  }
 \label{fig2}
  \end{figure*}
  All lengths are scaled by the bead diameter, ${\bf \tilde{r}} = {\bf r}/a$ and   energies by thermal energy, $u(\{ {\bf \tilde{r}}_N \}) k_B T = U(\{ {\bf     r}_N \})$ which leads to rescaled mobilities $ \bm{\tilde{\mu}}_{ij} =   \bm{\mu}_{ij} \Delta t k_B T/a^2$ with timestep $\Delta t$ and to a rescaled   shear rate $\dot{\tilde{\gamma}} = \dot{\gamma}a^2/(k_B T \mu_0 )$. The   random velocities which couple the system to a heat bath are modeled with   Gaussian white noise and fulfill the fluctuation dissipation theorem,   $\langle \bm{\tilde{\xi}}_i(t) \bm{\tilde{\xi}}_j(t') \rangle = 2   \bm{\tilde{\mu}}_{ij} \delta(t-t')$. 
The rescaled time step $\tilde{\mu}_0 =   \mu_0 \Delta t k_BT /a^2$ must be
 chosen small enough such that the bead displacement per time step is small compared to the bead radius.  
The total potential
$    u= u_\mathrm{surf} +   u_\mathrm{poly}$
 consists of bead-bead interactions
\begin{equation}
  \label{eq:intramol}
  u_\mathrm{poly} = k/2 \sum_i \left ( \tilde{r}_{ii+1} - 1 \right ) ^2
  + \epsilon \sum_{i<j} \left( \tilde{r}_{ij}^{-12} - 2 \tilde{r}_{ij}^{-6} \right) ,
\end{equation}
where $\tilde{r}_{ij}=\left| \tilde{\vect{r}}_j - \tilde{\vect{r}}_{i} \right|$ is the rescaled  monomer distance. 
The first term ensures the chain connectivity by harmonic bonds around the equilibrium length $a$ 
with a rescaled spring constant $k=200$, the second is a truncated Lennard-Jones potential with a rescaled 
parameter $\epsilon=2$ which is only used for $\tilde{r}_{ij}<1$ to avoid overlap of the chain.
For the attraction between surface and monomers an exponentially decaying potential is used,
\begin{equation}
  \label{eq:surface}
  u_\mathrm{surf} = - \chi \sum_i \expb{ -\tilde{\kappa} \tilde{z}_i } ,
\end{equation}
where $\tilde{\kappa}^{-1}=\kappa^{-1}/a$ and $\chi$ are the rescaled decay length and the rescaled interaction parameter, respectively. For electrostatically driven adsorption, the interaction parameter $\chi$ can be interpreted as the product of surface and polymer charge density and the decay length $\kappa^{-1}$ as the screening length. We use the exponential potential as a generic form to study short-ranged surface attraction and have in mind also hydrophobic or van-der-Waals interactions. To make it very short-ranged, we use $\tilde{\kappa}^{-1}=1/2$. We also include a hard wall interaction of the monomers at the wall-liquid boundary, \ie $u_\mathrm{surf}(\tilde{z}_i < 0)=\infty$.

We use a rescaled time step of $\tilde{\mu}_0=10^{-4}$ and simulate for at least 
$10^8$ simulation steps. The first $20\,\%$ of these steps are disregarded for equilibration. 
Every $10^3$ steps the configuration is recorded. Data are
obtained by block-averaging, statistical errors are obtained from the standard deviation of blocks and only shown when larger than the symbol size. Simulations are performed for polymers of typical  length $N = 64$ or $128$ 
and for different values of the adsorption strength ($\chi = 10^{-4} \ldots 10^2$).


We first briefly review the adsorption of a single polymer in equilibrium. 
In thermodynamic equilibrium, \emph{any} polymer of \emph{finite} length will desorb from
an adsorbing surface into the semi-infinite half-space, no matter how strong the adsorption potential is. 
However, by carefully choosing polymer length and simulation duration, information about polymer 
adsorption in the asymptotic limit of an infinitely long polymer can be obtained. 
This is so because a finite-time window exists within which 
an adsorbed polymer of finite length has enough time to equilibrate its conformation
but at the same time stays trapped in the surface adsorption potential. In fact, 
this time window widens with increasing monomer number $N$, or, conversely, shrinks for too short polymers.

The transition between the adsorbed and desorbed state can be quantified with different observables, 
\eg the adsorption potential energy \cite{Eisenriegler82}, the polymer mean height \cite{Yamakov99}
or  the number of adsorbed monomers \cite{Milchev96}. 
In fig. \ref{fig2}(a) we plot the mean adsorption potential per monomer, $- \langle {u}_\mathrm{surf}/ \chi N \rangle $, 
calculated from \eqnref{eq:surface}, 
for a 128mer as a function of $\chi$. 
In the thermodynamic limit $N \rightarrow \infty$ this observables is expected to go to zero in a 
continuous fashion at the adsorption transition. 


Next  we establish a simple linear response theory for the flattening of a chain in shear flow.
For simplicity, we consider a freely jointed chain (FJC) and neglect hydrodynamic interactions and
excluded-volume interaction between beads. We assume the chain to 
take some average separation from the surface.
From eq.~(\ref{langevin}) the x-component of the velocity of bead $i$ reads
$\dot{x}_i=\dot{\gamma}z_i + \mu_0 F^{(x)}_i- \mu_0 F^{(x)}_{i-1} + \xi_i^{(x)}$ where 
$F^{(x)}_i$ denotes the force that bead $i+1$ exerts on bead $i$ via 
the bonding potential. By reciprocity, $-F^{(x)}_{i-1}$ is the force that 
bead $i-1$ exerts on bead $i$ and the boundary condition 
reads as $F^{(x)}_N=F^{(x)}_0=0$. 
By averaging this equation, the random force disappears and we obtain 
$\langle \dot{x}_i \rangle =\dot{\gamma} \langle z_i\rangle  + \mu_0 \langle F^{(x)}_i \rangle 
- \mu_0 \langle F^{(x)}_{i-1} \rangle$. In the stationary state, all monomers have
the same average velocity equal to the mean chain velocity, 
$\langle \dot{x}_1 \rangle =  \langle \dot{x}_2 \rangle = \cdots$, which furnishes $N-1$ equations.
Solving for the forces, we obtain
\begin{equation} \label{force}
 \langle F^{(x)}_{i} \rangle     = -\frac{\dot{\gamma}}{\mu_0} \sum_{j=1}^i 
 \langle z_j - \sum_{k=1}^N z_k /N \rangle
 \end{equation}
which constitutes an exact relation between the average chain height profile and the mean 
forces acting on the polymer bonds. 

\begin{figure}
      \includegraphics[width=0.35 \textwidth]{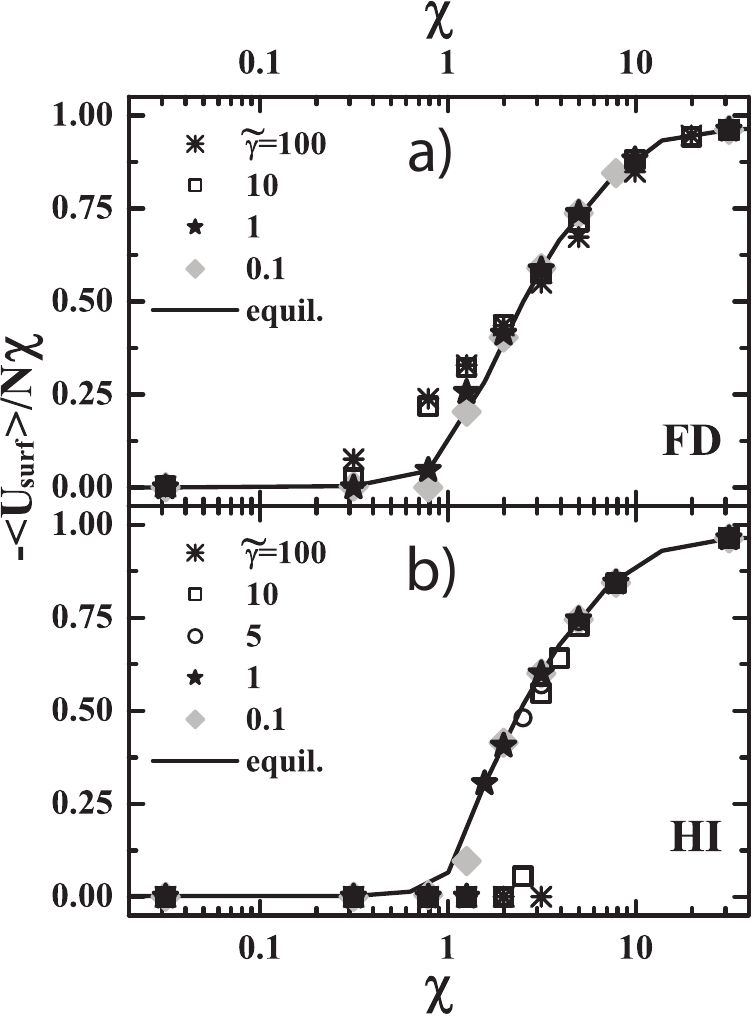}
    \caption{Mean normalized adsorption potential $ - \langle u_\mathrm{surf} \rangle/N\chi $ 
for a 64mer as a function of the surface interaction parameter $\chi$ in equilibrium (solid lines) and for various shear rates  
$\dot{\tilde{\gamma}}=\dot{\gamma} a^2 / \mu_0 \kBT$. 
a) Results from free-draining (FD) BD simulations  
and b) from hydrodynamic (HI) simulations.  In contrast to the FD the
HI simulations show weaker adsorption and a discontinuous desorption transition.}
 \label{fig3}
  \end{figure}

According to standard polymer statistical mechanics,
the mean extension of a FJC bond along the pulling direction is 
$\langle  \tilde{x}_{ii+1} \rangle  \equiv \langle  \tilde{x}_{i+1} - \tilde{x}_i \rangle =
\mathcal{L}(\tilde{F}_i)$ where 
$\mathcal{L}(x) = \coth{x} - x^{-1} $ is the Langevin function
and $\tilde{F}_i = a  \langle F^{(x)}_{i} \rangle / k_BT$ is a short-hand notation for the rescaled average
x-component of the bond force.
Likewise, the second moment is given by 
$\langle \tilde{x}_i^2 \rangle  = 1 - 2 \mathcal{L}(\tilde{F}_i)/\tilde{F}_i$\cite{serr}.
If we neglect effects of the surface, the two perpendicular directions, 
$\hat{y}$ and $\hat{z}$, are equivalent and the mean squared extension perpendicular to the pulling direction is
\begin{equation}
  \label{eq:fjc-zroughness}
  \avg{\tilde{z}_i^2} = \left(1 - \avg{\tilde{x}_i^2} \right) /2
  = \mathcal{L}(\tilde{F}_i) / \tilde{F}_i  
\end{equation}
with the asymptotic limits
$ \avg{\tilde{z}_i^2}   \stackrel{\tilde{F}_i  \ll 1}{\longrightarrow} 1/3$ for weak force and 
$ \avg{\tilde{z}_i^2}   \stackrel{\tilde{F}_i  \gg 1}{\longrightarrow} 1/ | \tilde{F}_i|  $ for strong force.
Since the root mean squared bond radius defines the Kuhn length, we can in the presence
of the lateral pulling force define an effective Kuhn length   in $\hat{z}$ direction  by 
$a^2_\mathrm{eff,i} =  3 a^2 \mathcal{L}(\tilde{F}_i) / \tilde{F}_i $. 
In fact, since the force varies along the chain contour,
it is useful to define the renormalized Kuhn length as an average over all bonds,
\begin{equation}
  \label{eq:kuhn-length}
  \bar{a}^2_\mathrm{eff} =  \frac{3 a^2 }{ N-1} \sum_{i=1}^{N-1} \frac{\mathcal{L}(\tilde{F}_i) }{ \tilde{F}_i} .
  \end{equation}

For illustration we show in \figref{fig1}(b-c)  the height profile and the force profile
at $\chi = 2$.
Due to the symmetry of the mean height profile
$\langle z_i \rangle$ of the chain and since  the chain ends have a larger separation from the surface than the chain middle,
according to eq.~(\ref{force}),
the force vanishes on the middle and terminal bonds and its magnitude is maximal in between.
Note that a non-zero bond force profile would even arise for e.g. polymer loops where by symmetry
the mean height profile is flat, caused by chain fluctuation around the mean (a mechanism
not pursued further in this paper).
Based on the mapping of the problem of a single polymer at an adsorbing surface
onto the equivalent problem of a quantum particle at an adsorbing wall\cite{Wiegel}, the only remaining
scaling variable in the problem  turns out to be the effective adsorption strength
\begin{equation} \label{chieff}
\chi_\mathrm{eff} = \chi  a^2 / \bar{a}_\mathrm{eff}^2.
  \end{equation}
This gives a direct clue as to how the chain flattening enhances chain adsorption.

  \begin{figure*}
    \begin{minipage}[t]{.32\textwidth}
      \includegraphics[width=\textwidth]{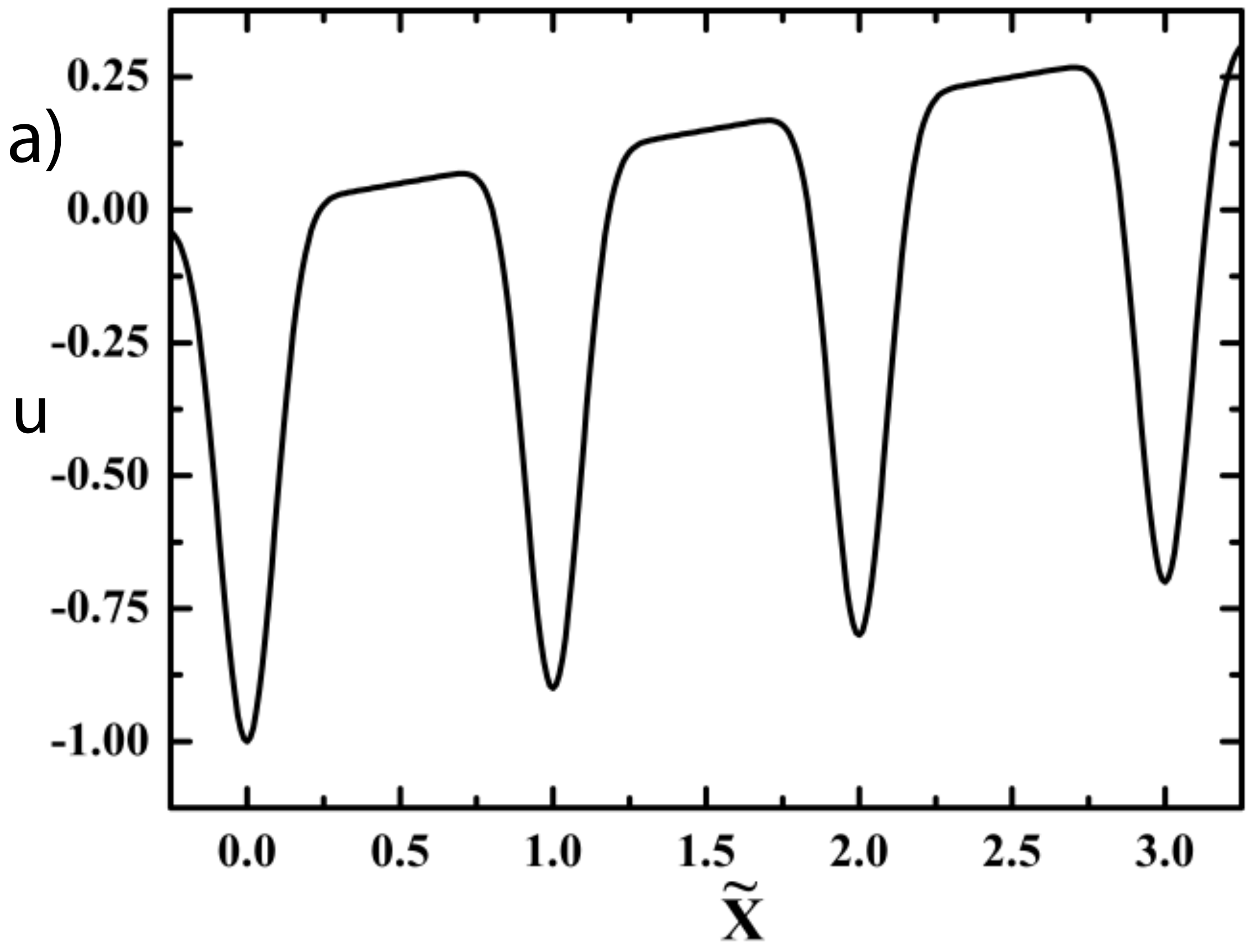}
    \end{minipage}
    \hfill
    \begin{minipage}[t]{.32\textwidth}
      \includegraphics[width=\textwidth]{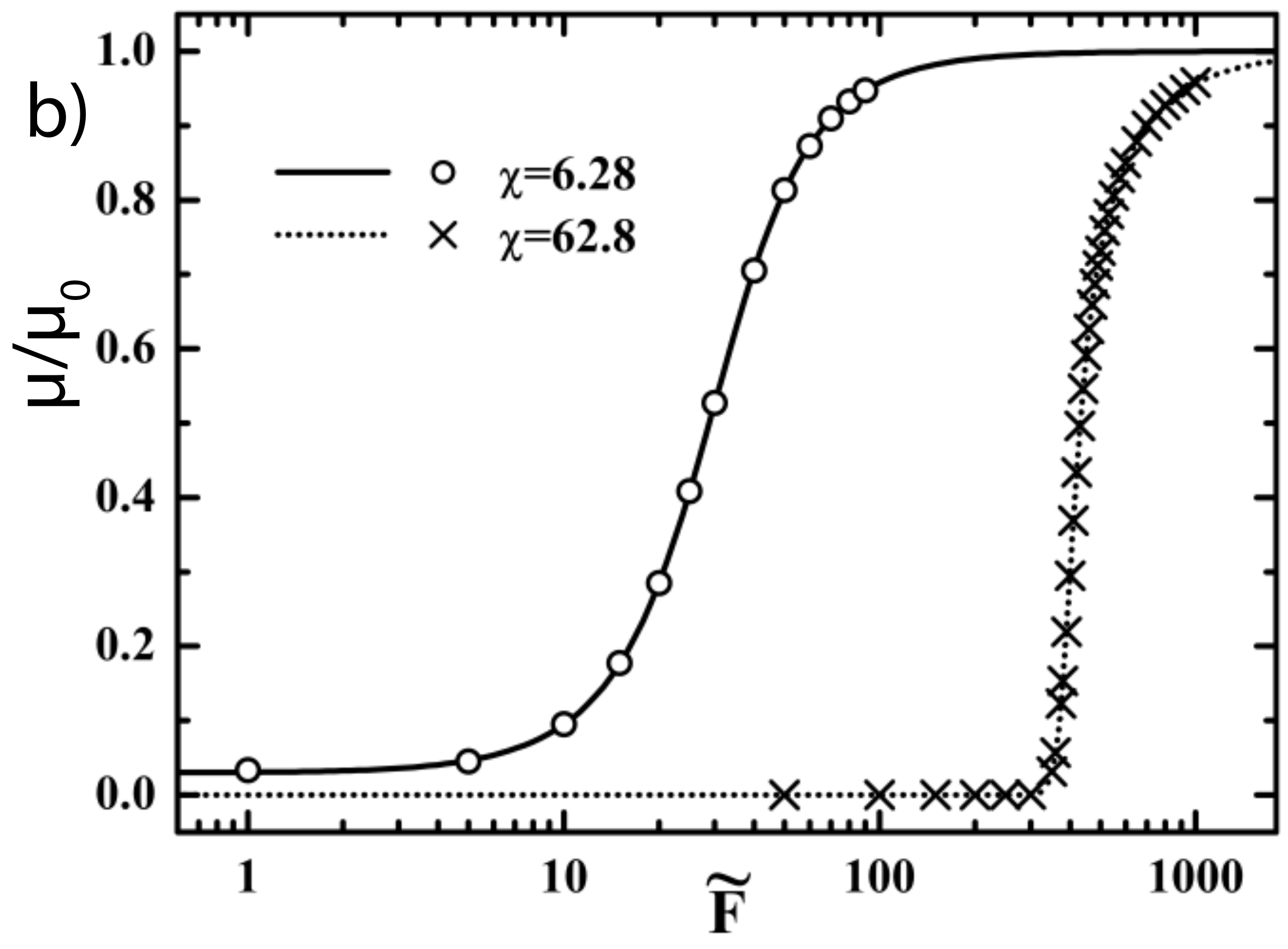}
    \end{minipage}
   \hfill
    \begin{minipage}[t]{.32\textwidth}
      \includegraphics[width=\textwidth]{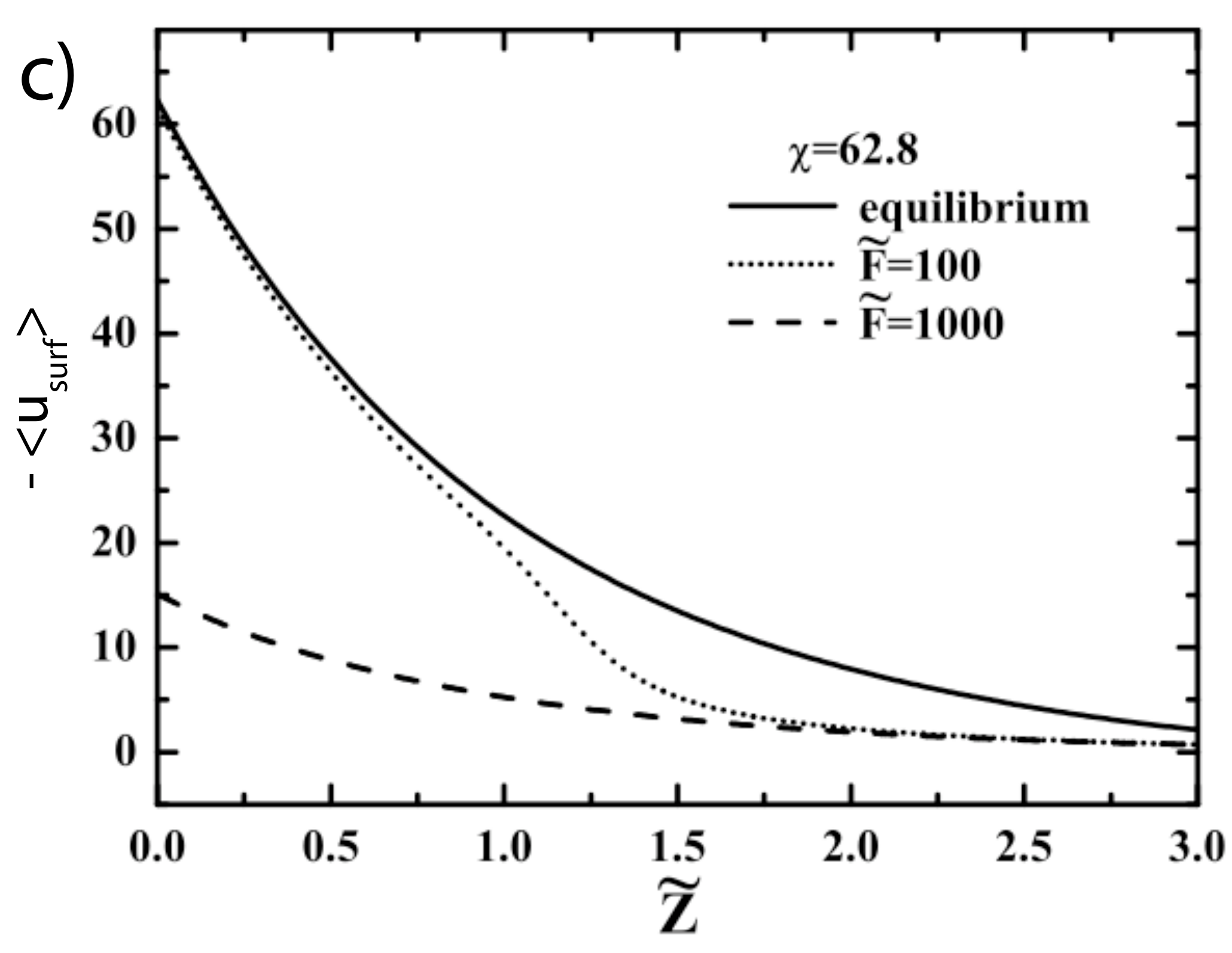}
    \end{minipage}
    \caption{(a) Potential energy $u$  
 of a single particle at fixed height $\tilde{z}=0$ in a surface potential \eqnref{co} 
 and pulled with a rescaled force of $\tilde{F}=-0.1$. The surface interaction strength is chosen as $\chi=1$, 
 corrugation power is $\beta=6$.
 (b) Average mobilities $\mu/\mu_0$ for a single particle pulled through a 
corrugated potential eq.~(\ref{co}) as a function of the rescaled pulling force 
$\tilde{F}=Fa /k_BT$ for fixed  particle height $\tilde{z}=0$,  $\beta=6$, 
$\tilde{\kappa}=1$, $\chi=6.28$ and $\chi=62.8$. 
Data points are taken from simulations, the lines from Fokker-Planck theory
eq.~(\ref{FPsol}). 
(c) Average surface potential $-\langle u_\mathrm{surf} \rangle $ for a single particle as a 
function of its surface separation  $\tilde{z}$ from  stationary solutions of the Fokker-Planck eq.~(\ref{FPsol}) 
with surface potential parameters $\tilde{\kappa}=1$, $\beta = 6$, and $\chi=62.8$. 
For intermediate force $\tilde{F}=100$  the average potential crosses over from the
equilibrium limit ($\tilde{F}=0$, solid line) to the high-force limit ($\tilde{F}=1000$, dashed line)
at intermediate separation.
 }
 \label{fig4}
  \end{figure*}

As exemplary case we show in fig. \ref{fig2}(a) the mean normalized adsorption potential 
$ -\langle u_\mathrm{surf}\rangle / (N\chi) $ for a 128mer  from free-draining simulations
as a function of the adsorption strength 
$\chi$ for the equilibrium case (solid line) and for different  shear rates $\dot{\tilde{\gamma}}$ (symbols). 
For low values of the adsorption strength $\chi$ we note 
shear-enhanced adsorption in agreement with previous free-draining simulations\cite{Manias95,Panwar05}. 
At large values of $\chi$ the  adsorption behaviour is not modified. 
Intuitively, one might expect peeling and tumbling effects to reduce adsorption at
strong shearing, but nothing of this sort is seen.
In   fig. \ref{fig2}c we show the fraction of adsorbed monomers or trains, 
$\langle \Upsilon \rangle /N$, where a monomer counts as adsorbed when its height is less than the 
adsorption screening length, $\tilde{z}_i< \tilde{\kappa}^{-1}$ in the strong adsorption regime.
Almost no influence of shear is seen.
This is easily understood when taking the effects of chain flattening into account. 
For high adsorption strengths, the polymer adopts a flat configuration with long trains and only few, small loops.
 The stretching forces acting on the polymer bonds are low and chain  flattening is negligible
 and thus  the enhanced adsorption mechanism discussed previously does not come into play. 
 On the other hand, close to the desorption transition loops proliferate  
 and experience stronger shear forces,  which leads to chain flattening and thus
enhanced polymer adsorption.
On the linear-response level, the configuration and adsorption energy of a chain in shear can be approximated
by the equilibrium configuration of a chain  (i.e. without shear) with the adsorption strength
$\chi$ replaced by $\chi_\mathrm{eff} $ given by eq.~(\ref{chieff}). 
The result of this procedure is shown in fig \ref{fig2}(b). 
The linear prediction qualitatively captures the shear-induced adsorption enhancement but
grossly  overestimates the effect.
In fact, linear response breaks down at strong shear flows since the resulting chain
 flattening of the chain very efficiently decreases the shear effects, this is even more enhanced
 by the stronger adsorption of the flattened chain. The linear theory can in principle be self-consistently improved,
 which we do not pursure
since in the following we show that hydrodynamic lift effects dominate chain flattening.

In fig. \ref{fig3} we plot the rescaled adsorption potential 
$-\langle u_\mathrm{surf} \rangle  /N\chi$ 
as a function of the adsorption strength $\chi$ for the equilibrium case and various shear rates for 
chains of length $N=64$. 
In a) we show the results from free-draining (FD) simulations, in b) 
we show data for the exact same parameters  using full hydrodynamic interactions (HI). 
As expected, the equilibrium results (solid lines) from both methods coincide within the precision of the simulations. 
For moderate shear rates the HI simulations quantitatively confirm the results obtained in the FD case. 
For elevated shear rates, however,  the transition into the desorbed state changes from continuous (FD) to discontinuous (HI)
and the magnitude of the adsorption energy is smaller compared to the equilibrium case. We conclude
that hydrodynamic effects weaken adsorption in shear and totally dominate the chain flattening effects
seen in the absence of correct hydrodynamic effects. 

What is the cause for this drastic hydrodynamic repulsion? 
\revision{It is known that dumbbells\cite{Graham} as well as stiff rods and flexible polymers\cite{Ladd,sendner} 
close to a flat, non-adsorbing surface
in shear experience a repulsive lift force away from the surface. 
This lift force can be rationalzed by the anisotropy of the hydrodynamic mobility parallel and normal to an elongated object\cite{doi} 
in conjunction with  the anisotropically distributed orientation of the object in shear \cite{sendner}. 
The resulting  potential is long-ranged and decays as
$\propto 1/z$  for rods as well as  flexible polymers at elevated shear rates\cite{sendner}.}
The combination of such a long-ranged repulsive potential with a short-ranged attractive surface
adsorption potential is known to turn  the adsorption transition of a polymer discontinuous\cite{Lipowsky}. 
Our simulations confirm the predicted change in the nature of the transition.
In summary, due to the relative weak enhancement of adsorption due to chain flattening observed
in free-draining simulations, the hydrodynamic lift repulsion dominates and weakens the adsorption
and actually turns the transition discontinuous.

Up to now the adsorbing wall was assumed  flat and homogeneous. 
This is a good approximation for atomically flat surfaces 
or adsorption driven by hydrophobic interactions since here the surface friction effects are
small \cite{Kuhner,serrJACS}.
Also for charged substrates the assumption of a laterally homogeneous potential is accurate
 if the screening length is larger than the 
 monomer bond length and the separation between charged sites on the substrate. 
 In all other cases, and in specific for adsorption driven by hydrogen-bonds\cite{serrJACS},
 adsorbing polymers do experience corrugated potentials the non-equilbrium effects of which
 will be briefly discussed now. Since the equilibration of a polymer in a corrugated potential landscape
 including hydrodynamic interactions is by far too demanding from the computer time aspect, we
 consider the simplistic case of a single monomer dragged by an external force 
in a one-dimensional corrugated potential.
 As a straightforward extension of the potential  eq.~(\ref{eq:surface}) used in the simulations, we consider
 the total potential $ u (\tilde{x}, \tilde{z})= u_\mathrm{surf} (\tilde{x}, \tilde{z}) - \tilde{F} \tilde{x}$
 including the corrugated surface potential
 \begin{equation}
  u_\mathrm{surf} (\tilde{x}, \tilde{z}) = - \chi \expb{ -\tilde{\kappa} \tilde{z}}
  \left [ \frac { \cos \left ( 2\pi \tilde{x} \right) + 1 } {2} \right] ^{\beta} 
  \label{co}
\end{equation}
where the exponent $\beta$ controls the corrugation steepness,
the corrugation wavelength is for simplicity set equal to the monomer diameter $a$,
and the force $\tilde{F}$ leads to particle motion. 
Here we concentrate on the highly
corrugated case $\beta=6$ shown in fig. \ref{fig4}(a) for $\tilde{z}=0$, $\tilde{F}=-0.1$ 
and $\chi=1$.
We consider 
numerical solutions of the stationary Fokker-Planck (FP) equation\cite{risken}
$ 0=  \dot{P}(\tilde{x},\tilde{t}) = u'(\tilde{x}) P'(\tilde{x},\tilde{t})
+u''(\tilde{x}) P(\tilde{x},\tilde{t}) + P''(\tilde{x},\tilde{t})$,
where $P(\tilde{x},\tilde{t})$ is the normalized distribution function,
 the rescaled time is $\tilde{t} = t \mu_0 k_BT / a^2$
 and $u'(\tilde{x}) = 
 \partial u(\tilde{x}) / \partial \tilde{x}$
denotes a partial derivative w.r.t. $\tilde{x}$ (and similarly for $P$). 
Note that $\tilde{z}$ only enters parametrically into the FP equation,
i.e., the particle is for illustrative purposes constrained to fixed height $\tilde{z}$
and we therefore dropped the $\tilde{z}$-dependence of the potential.
For time-independent potentials consisting of a linear and a part periodic over the interval 
from $\tilde{x}=0$ to $\tilde{x}=1$, 
the stationary solution is given by\cite{risken}
\begin{equation}
  P(\tilde{x}) = \frac{ e^{-u(\tilde{x})} } {Q} 
    \left[ \frac{  \int_{0}^{1}d \tilde{x}'{  e^{u(\tilde{x}')}  } }
      {1 - e^{u(1) - u(0)} }
    - \int_{0}^{\tilde{x}} d \tilde{x}' e^{u(\tilde{x}')} \right ] 
  \label{FPsol}
\end{equation}
where $Q$ is the normalization constant. 
The rescaled mobility follows  as
$  \mu / \mu_0 = \langle \dot{\tilde{x}} \rangle / (\tilde{F} \tilde{\mu}_0)
 =-  \int_{0}^{1}d \tilde{x}  u'(\tilde{x})  P(\tilde{x})  / (\tilde{F} Q) $
 and the average potential energy follows as 
$  \langle u_\mathrm{surf} \rangle  =  
\int_{0}^{1}d \tilde{x}  u_\mathrm{surf}(\tilde{x})  P(\tilde{x})  / Q $.
\revision{ In fig. \ref{fig4}(b) the normalized 
particle mobility $\mu/\mu_0$ from  the FP solution (shown as lines)
compares not surprisingly  favorably with the Brownian dynamics simulation results (data points)
according to eqs.~(\ref{FD}) and (\ref{eq:5})
at fixed particle height $\tilde{z}=0$.}
Data is presented for $\chi = 6.28$ (circles) and $\chi = 62.8$ (crosses). 
For low external force, the particle mostly  sticks to the attractive surface sites 
and the average mobility is very small. 
At large forces the average mobility approaches its bulk value and the 
motion is uniform regardless of the potential corrugation.
In  fig. \ref{fig4}(c) we show the average potential energy 
$ - \langle u_\mathrm{surf} \rangle$ as a function of the vertical height $\tilde{z}$ 
for fixed surface interaction parameter $\chi=62.8$ and for different  driving forces.
For comparison, the equilibrium case (i.e. for vanishing force) is shown as a solid line. 
For the larger force  $\tilde{F}=1000$ the mobility is not modified (as seen from fig. \ref{fig4}(b) for 
$\tilde{z}=0$)  by the 
potential and the particle moves uniformly over the surface, the average potential is therefore a uniform 
average over the corrugation. In contrast, in equilibrium, the particle mostly samples
the potential minima and the magnitude of the mean potential is much larger.
For intermediate force  $\tilde{F}=100$ an interesting phenomenon appears: 
For small values of $\tilde{z}$ the
particle sticks to the potential minima and the average potential is 
close to  the equilibrium case. With increasing height $\tilde{z}$ the lateral
confinement due to the potential is weakened, the mobility increases, and the average
potential crosses over to the high-force limit.
This means that the effective potential in the presence of a lateral driving  force
becomes more short-ranged, under the influence of shear where the driving force increases
with distance from the surface this effects will be even more pronounced.

For  polymer adsorption in shear 
this means  that potential corrugations  will in the presence of lateral force or shear
 tend to weaken
the attraction when compared to the equilibrium case.
Neglecting the enhanced friction close to the surface,
which could lead to chain flattening effects,
our tentative conclusion is that potential corrugation
will act similarly as hydrodynamic lift effects and 
favor polymer desorption,
\revision{ in line with simulation results for charged polymers\cite{Kumar3}}. 
Furthermore,  since the effective corrugated  surface potential becomes 
more short-ranged compared to the equilibrium case,
the desorption transition should be even more discontinuous compared to a homogeneous surface
potential.

Interesting effects left out in the present study include multi-chain effects, chain stiffness effects
and structural effects that go beyond modelling  chain monomers as a homogenous spheres. 

\acknowledgements
We acknowledge  support by the Elitenetzwerk Bayern in the framework of CompInt
and by the DFG (NE 810-4).

\end{document}